\documentclass{article}
\pdfoutput=1
\usepackage[english]{babel}
\usepackage[margin=1.3in]{geometry}
\usepackage{multicol}
\usepackage{background}
\usepackage{caption}

\SetBgScale{4}
\SetBgColor{gray}
\SetBgAngle{90}
\SetBgContents{}
\SetBgPosition{-1.5,-10}
\begin{document}

{\centering
  
  {\bfseries\Large Combined Opto-Acoustical Sensor Modules for KM3NeT\bigskip}
  
  A.~Enzenh\"ofer on behalf of the KM3NeT consortium \\
  {\it{Friedrich-Alexander-Universit\"at Erlangen-N\"urnberg, Erlangen Centre for Astroparticle Physics, Erwin-Rommel-Str.\,1, D-91058 Erlangen, Germany}} \\
    \Large August 19, 2014
    
}

\begin{abstract}
KM3NeT is a future multi-cubic-kilometre water Cherenkov neutrino telescope currently entering a first construction phase.
It will be located in the Mediterranean Sea and comprise about 600 vertical structures called detection units.
Each of these detection units has a length of several hundred metres and is anchored to the sea bed on one side and held taut by a buoy on the other side.
The detection units are thus subject to permanent movement due to sea currents.
Modules holding photosensors and additional equipment are equally distributed along the detection units.
The relative positions of the photosensors has to be known with an uncertainty below 20\,cm in order to achieve the necessary precision for neutrino astronomy.
These positions can be determined with an acoustic positioning system: dedicated acoustic emitters located at known positions and acoustic receivers along each detection unit.
This article describes the approach to combine an acoustic receiver with the photosensors inside one detection module using a common power supply and data readout.
The advantage of this approach lies in a reduction of underwater connectors and module configurations as well as in the compactification of the detection units integrating the auxiliary devices necessary for their successful operation.
\end{abstract}

\section{Introduction}

The future neutrino telescope KM3NeT \cite{km3net} will instrument several cubic kilometres of water in the deep Mediterranean Sea.
The detection principle is based on the detection of the faint Cherenkov light emitted by charged particles which are created in neutrino-nucleon interactions.
This light is detected by photosensors consisting of PMTs and subsequent readout electronics.
About 20 modules of photosensors will be integrated into vertical structures called detection units (DUs) with equal spacing.
600 DUs are required to satisfactorily instrument the intended detection volume.
The detection units have a length of several hundred metres and will be anchored to the sea bed on one side and held taut by a buoy on the other side.
In this configuration each detection unit and thus each photosensor is subject to permanent movements due to sea currents.
In order to achieve a nanosecond timing resolution for the photosensor hits required for the precise event reconstruction it is necessary to determine the relative positions of these photosensors with a precision of better than 20\,cm.
A common way to obtain these positions with the necessary accuracy is the use of an acoustic positioning system.
There are several commercial systems for this task but their application is mainly restricted to this specific task.
In order to take advantage of the infrastructure provided by KM3NeT it is desirable to maximise the use of the acquired acoustic data by including additional objectives like monitoring of the deep sea acoustic environment or studies towards acoustic detection of high-energy neutrino interactions in the sea water.
Several activities are ongoing to develop new sensor systems capable of performing these tasks.

This article gives a short introduction to the acoustic positioning system and focuses on a new special type of receiver, the so called Opto-Acoustical Module developed at ECAP\footnote{Erlangen Centre for Astroparticle Physics} \cite{ECAP}.

\section{Acoustic positioning system}

\begin{figure}
  \begin{center}
    \includegraphics[width=0.5\linewidth]{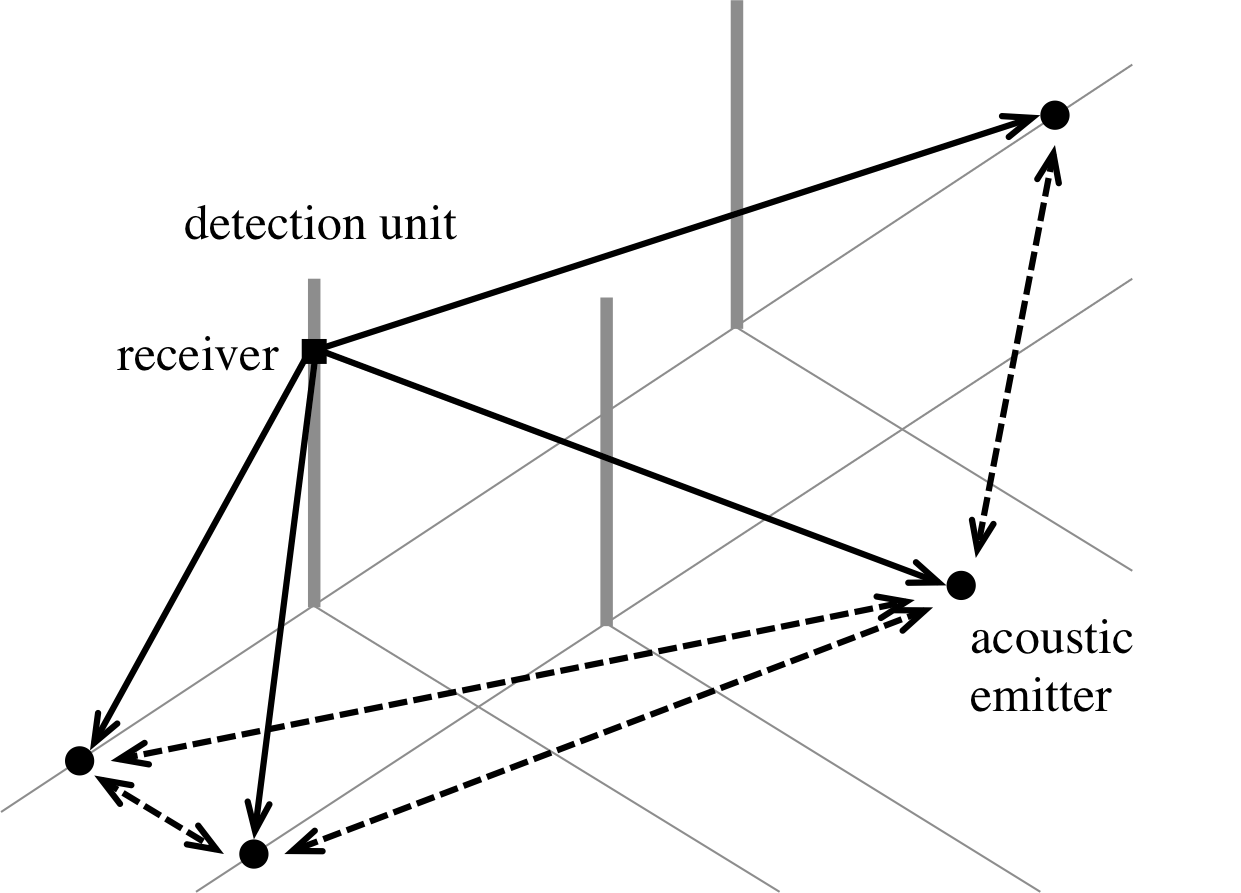}
    \caption{Schematic of an acoustic positioning system showing one receiver on a detection unit and several acoustic emitters on the sea bed. The dashed lines depict known distances between the emitters, whereas the other lines correspond to the calculated distances between the receiver and each of the emitters.}
    \label{fig_aps}
  \end{center}
\end{figure}

The acoustic positioning system consists of several acoustic emitters located at known positions and receivers attached to the moving parts of the detector.
The emitters will be located near the sea bed with a power and signal connection to the DU.
It is possible to emit different signal shapes with variable amplitude and timing.
The acoustic receivers will be spread throughout the detector on defined positions in order to facilitate the reconstruction of the relative positions of the DUs, cf.~Fig.~\ref{fig_aps}.
The active parts of the acoustic positioning system employ the piezoelectric effect.
This effect provides stable and reliable operation of transceivers and sensors, also under the high ambient pressure of the deep sea.
The protection against the chemically aggressive sea water can be performed mainly in two ways.
Firstly it is possible to cast the devices in adequate inert material providing adapted sound transmittivity, like polymers.
Secondly, the devices can be installed inside a pressure resistant housing, e.g.~made of glass as for the photosensors.
Through subsequent emission and reception of sound signals from the individual emitters it is possible to determine the positions of all receivers relative to the emitters:
\begin{equation}
\label{equ_1}
\left|\vec{r}_{reception} - \vec{r}_{emission} \right| = c_{sound} \times \left( t_{reception} - t_{emission} \right)
\end{equation}
Using trilateration, these distances finally yield the relative positions.

\section{Opto-Acoustical Modules}

Underwater connectors constitute vulnerabilities of the mechanical structures which should be reduced to a minimum.
The use of combined devices is a way to achieve this without reducing the number of sensors.
The combined device described here will combine optical sensors with acoustical sensors to so-called ``Opto-Acoustical Modules'' (OAMs).
The optical sensors can be any kind of sensors used for the detection of Cherenkov photons.
There is no limitation on the specific PMT type as well as on the number of PMTs.
The relatively small cylindrical dimensions of the acoustical sensor of $21\,\textrm{mm} \times 30 \, \textrm{mm}$ allows for its integration into a variety of different module designs.
These designs include modules with single large-area photomultiplier tubes (e.g.~in NEMO Phase \MakeUppercase{\romannumeral 2} \cite{NEMO}) or modules with several small photomultiplier tubes (e.g.~the KM3NeT multi-PMT design \cite{km3net_tdr}).

The number of individual acoustical sensors per module depends on the angular acceptance of the sensors as well as on their intended use.
A study of the angular acceptance of a piezo glued to an empty glass sphere can be found in \cite{diplom_ae}.
The acoustic positioning of the DUs only requires a single acoustical sensor per module as the preferred direction of incoming sound waves is given by the fixed positions of the acoustic emitters.

\begin{figure}
  \begin{center}
    \includegraphics[width=0.4\linewidth]{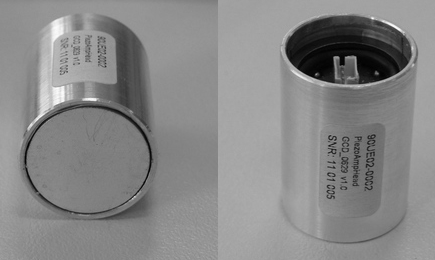}
    \caption{Front view (left) and side view (right) of the sensor unit, the front face of which is directly glued to the glass sphere.}
    \label{fig_sensor_head}
  \end{center}
\end{figure}

\begin{figure}
  \begin{center}
    \includegraphics[width=0.5\linewidth]{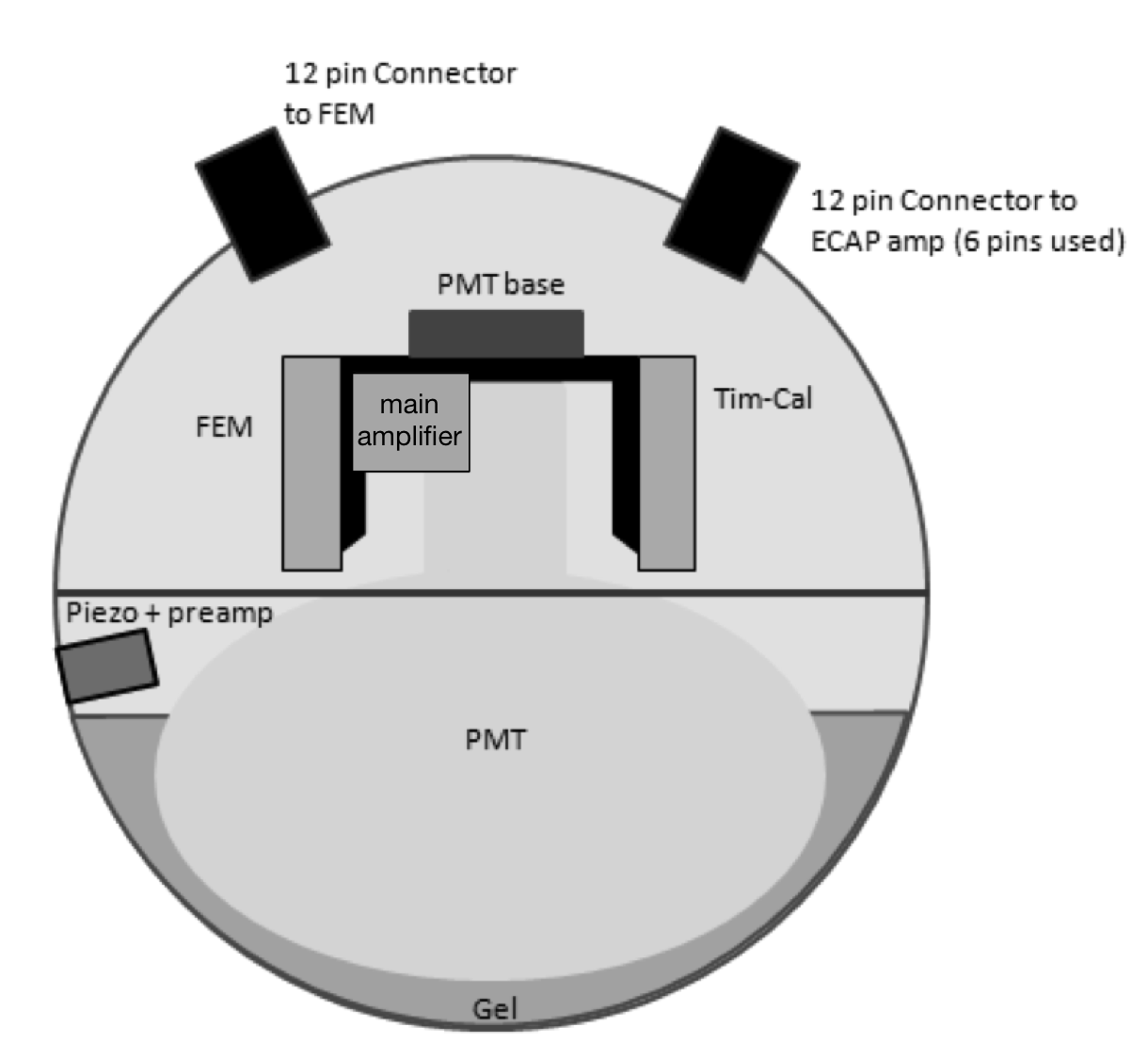}
    \caption{Schematic of a OAM prototype which will be integrated into NEMO Phase \MakeUppercase{\romannumeral 2}. The acoustic sensor is shown as piezo + preamp(lifier) on the left and the main amplifier board next to the PMT base. The FEM (Front End Module) is used to control the PMT, i.e.~setting and monitoring of the applied high voltage and other parameters. Tim-Cal is an internal device to check the time calibration of the PMT.}
    \label{fig_nemo_scheme}
  \end{center}
\end{figure}

The acoustical sensor itself consists of a cylindrical piezo ceramic (diameter\,18\,mm, height 12\,mm) connected to a pre\-amplifier; both are integrated inside an aluminum tube of $21\,\textrm{mm} \times 30\,\textrm{mm}$.
The aluminum tube serves as common ground for the acoustical sensor and ensures sufficient heat conduction to the glass sphere and thus to the deep sea environment.
This part of the acoustical sensor (cf.~Fig.~\ref{fig_sensor_head}) is glued directly into the glass sphere with a rigid 2-component adhesive on a epoxy resin base to ensure a good coupling to the glass.
In addition, a main amplifier board housing a bandpass filter ($2\,\textrm{kHz} - 125\,\textrm{kHz}$) conditions the data for the transport via LVDS\footnote{Low Voltage Differential Signaling} to the ADC which is part of the readout electronics of the module.
The main amplifier board is a single-layer PCB with a dimension of \mbox{$48\,\textrm{mm} \times 32\,\textrm{mm}$}.
This board can be integrated anywhere in the glass sphere, preferably next to the readout electronics of the module.
The whole acoustic sensor is designed for low power consumption ($< 40\,\textrm{mA}\,@\,3.3\,\textrm{V}$) to simplify its integration into already existing infrastructures.
Its easy integration was proven in two different designs.
Two prototype modules where build with a single-PMT design which is schematically shown in Fig.~\ref{fig_nemo_scheme} and which will be integrated into NEMO Phase \MakeUppercase{\romannumeral 2}.
These modules consist of a $13\,''$ sphere housing a $10\,''$ PMT.
Each prototype module is converted to an OAM by just including the acoustic sensor inside the sphere.
Due to its small dimensions, the acoustic sensor can be placed right above the optical gel as can be seen in Fig.~\ref{fig_nemo_scheme}.

\begin{figure}
  \begin{center}
    \includegraphics[width=0.7\linewidth]{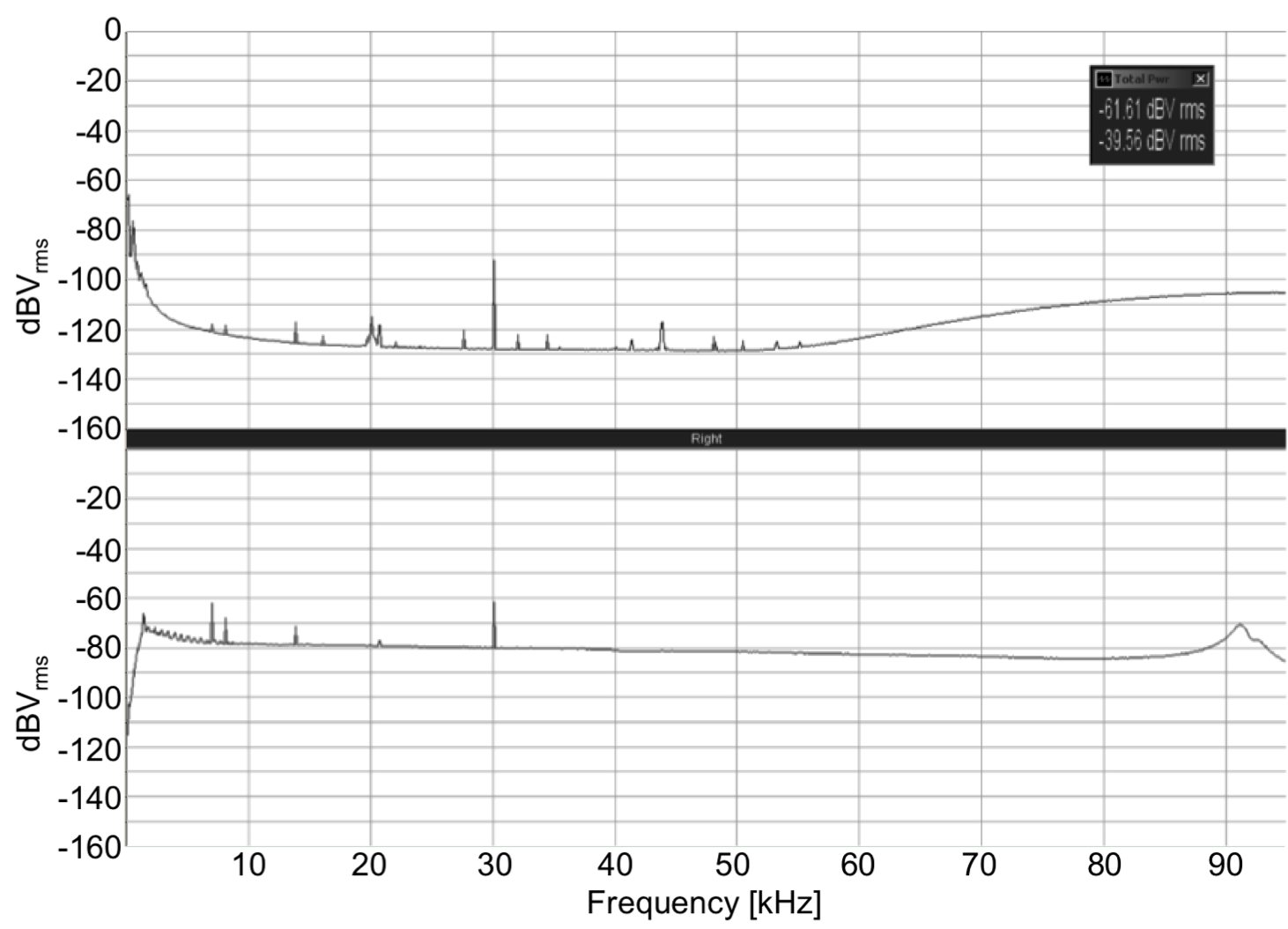}
    \caption{Comparison of frequency spectra for a SMID hydrophone (top) and an acoustic sensor as part of the OAM (bottom). A sinusoidal signal with a frequency of $ 30\,\textrm{kHz}$ was applied. The figure is further described in the text.}
    \label{fig_nemo_spectrum_with_signal}
  \end{center}
\end{figure}

\begin{figure}
  \begin{center}
    \includegraphics[width=0.7\linewidth]{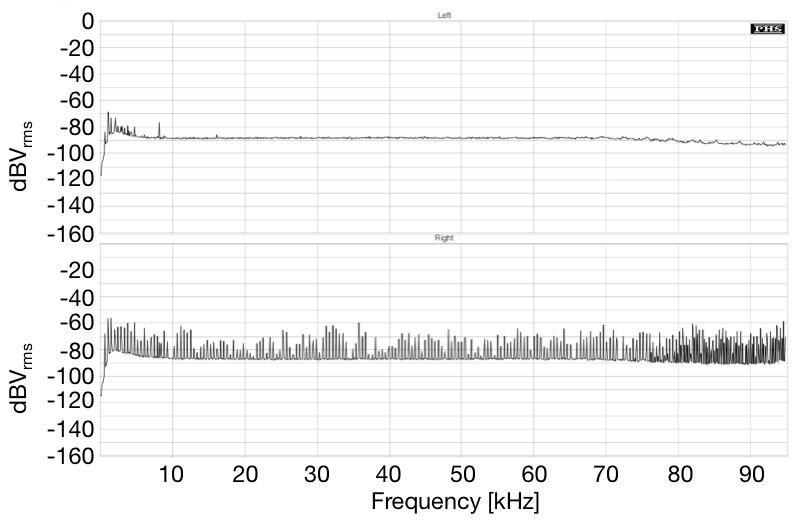}
    \caption{Comparison of frequency spectra measured in the laboratory with the Tim-Cal board inactive (top) and active (bottom). Its influence on the acoustic sensor is clearly visible and will be investigated for further developments.}
    \label{fig_nemo_spectrum_with_timcal}
  \end{center}
\end{figure}

There was no need to adapt the power supply and the data readout system to the acoustic sensor.
The first tests of this prototypes in cooperation with \mbox{INFN-LNS} in Catania \cite{lns_infn} confirm the feasibility to operate this kind of device.
Figure~\ref{fig_nemo_spectrum_with_signal} depicts the frequency spectrum measured in air in comparison to that of a SMID \cite{smid} hydrophone.
This type of hydrophone will be used within NEMO Phase \MakeUppercase{\romannumeral 2} as standard receiver for the acoustic positioning system.
The sinusoidal ultrasonic signal with a frequency of $30\,\textrm{kHz}$ is clearly visible in both spectra besides some additional spikes.
The SMID hydrophone has a very low noise level in the frequency range \-between $5\,\textrm{kHz}$ and $55\,\textrm{kHz}$.
The frequency spectrum of the acoustic sensor is essentially flat in the whole frequency range from $5\,\textrm{kHz}$ and $85\,\textrm{kHz}$.
Around $92\,\textrm{kHz}$ the piezo resonance is visible in this configuration.
The operation of the PMT next to the acoustic sensor resulted in an increase of the electromagnetic noise measured by the acoustic sensor.
A drawback was observed when switching on a specific calibration device (Tim-Cal) inside the OAM (cf.~Fig.~\ref{fig_nemo_spectrum_with_timcal}).
This leads to a large amount of electro\-magnetic interference with the acoustic sensor.
This is, however, just a minor inconvenience as this device is only used occasionally for calibration purposes and not in regular operation.
Moreover, this interference can be reduced with a design adapted to this problem.

\begin{figure}
  \begin{center}
    \includegraphics[width=0.5\linewidth]{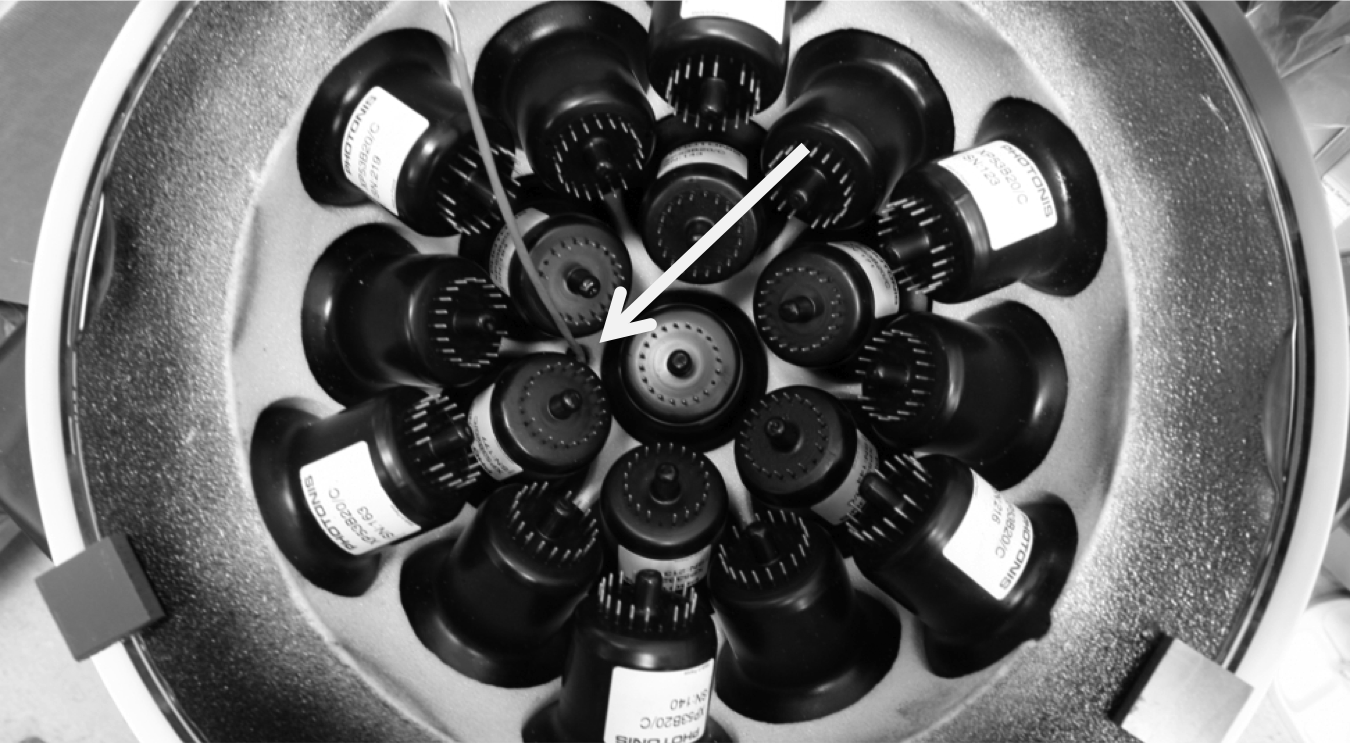}
    \caption{The PMTs in this design are tightly packed inside the hemispheres and only allow for small additional sensors. The cable running to the top left belongs to the acoustic sensor (position indicated by an arrow).}
    \label{fig_nikhef_inside}
  \end{center}
\end{figure}

\begin{figure}
  \begin{center}
    \includegraphics[width=0.4\linewidth]{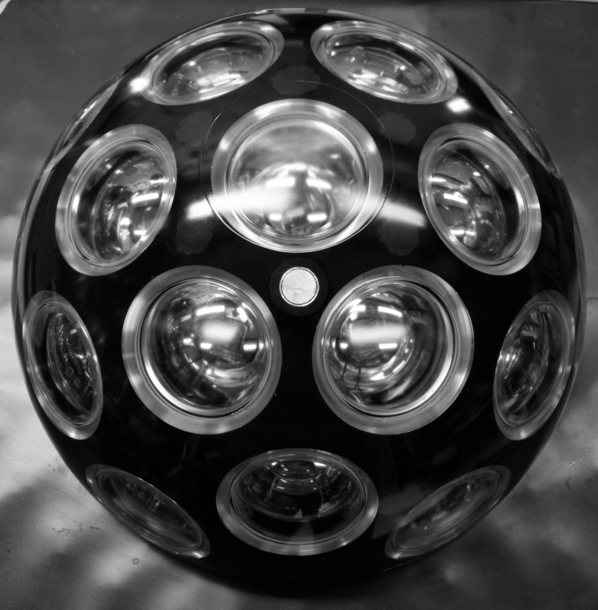}
    \caption{The small spot in the middle of the picture is one of the metallised piezo electrodes. This design confirms the good integrability of the acoustic sensor in complex optical module designs.}
    \label{fig_nikhef_outside}
  \end{center}
\end{figure}

Another prototype module with a different design was manufactured in cooperation with Nikhef \cite{nikhef}.
This multi-PMT module houses 31 $3\,''$ PMTs which are held by a foam structure to ensure their correct positions inside both hemispheres, cf.~Fig.~\ref{fig_nikhef_inside}.
The foam structure offers some small regions between the PMTs to include one or more acoustic sensors in this modules as can be seen in Fig.~\ref{fig_nikhef_outside}.

\begin{figure}
  \begin{center}
    \includegraphics[width=0.5\linewidth]{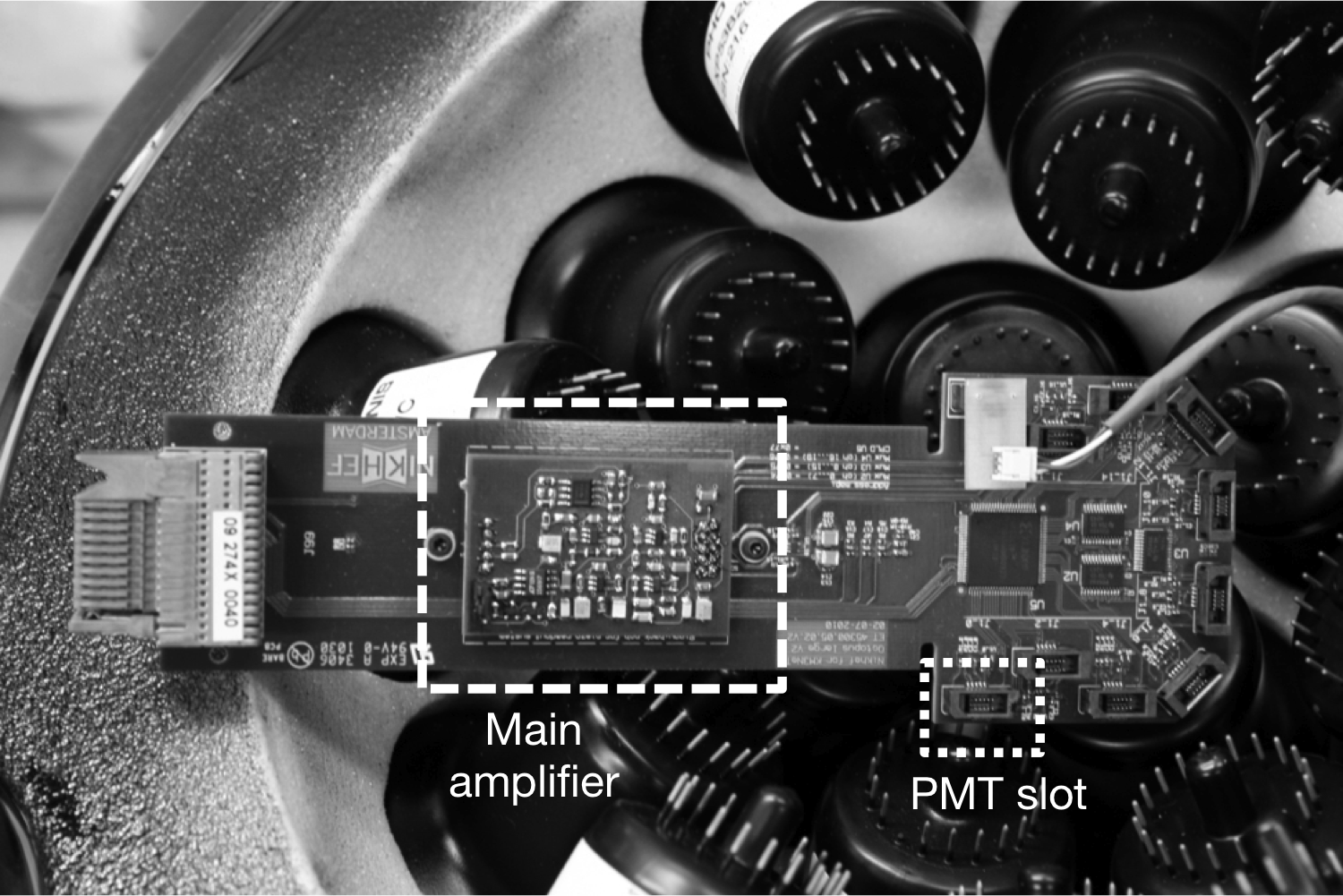}
    \caption{The main amplifier board attached to the centre of the octopus board. On the right-hand side the small cable visible in Fig.~\ref{fig_nikhef_inside} is connected to the octopus board via a small adapter. None of the PMTs is connected so their slots on the board stay empty.}
    \label{fig_nikhef_connector}
    \end{center}
\end{figure}

Figure~\ref{fig_nikhef_connector} shows the main amplifier board attached to the so-called ``octopus'' board providing signal connection and power supply to all sensors of the module.
In this case the main amplifier board was designed to match the specifications of the octopus board, so it was possible to reserve a dedicated space to simplify its integration.

\begin{figure}
  \begin{center}
    \includegraphics[width=0.7\linewidth]{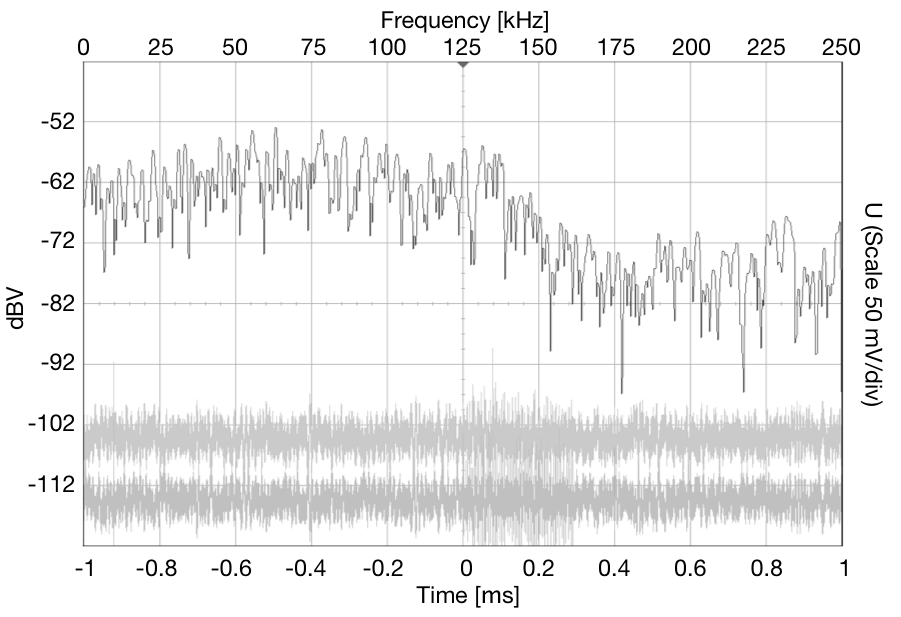}
    \caption{Noise spectrum of the acoustic sensor (upper part). The sensor and five PMT bases (without PMT) were powered through the octopus board. At about $125\,\textrm{kHz}$ (in the centre of the spectrum) the filter characteristics become clearly visible. The two differential LVDS signals are also present at the bottom of this figure. No averaging was applied to the recorded signals.}
    \label{fig_nikhef_spectrum}
  \end{center}
\end{figure}

\begin{figure}
  \begin{center}
    \includegraphics[width=0.7\linewidth]{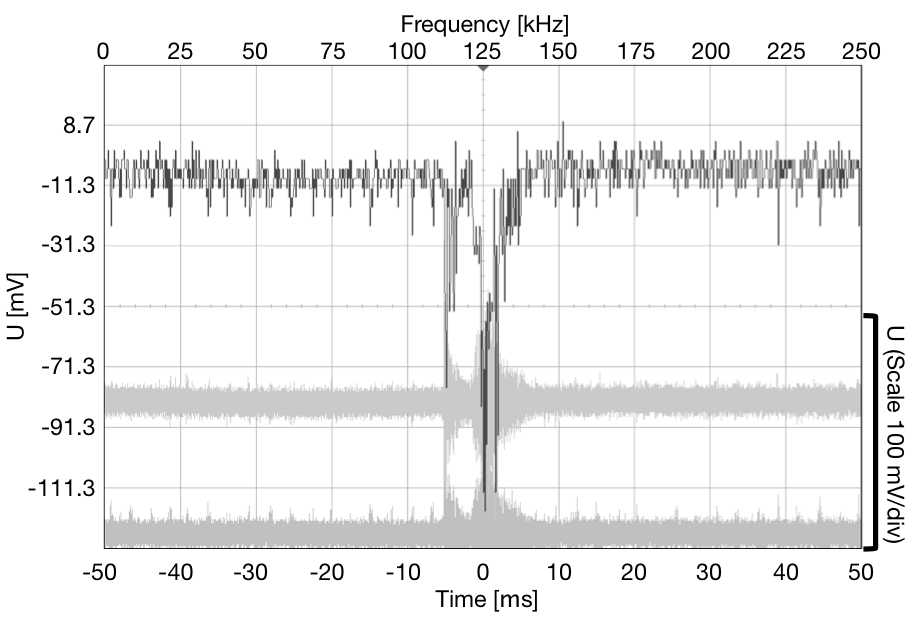}
    \caption{Result of a functional check of the acoustic sensor by tapping on the glass sphere with a fingertip. One of the differential signals at the bottom is partially out of the range of the oscilloscope. This leads to the distorted signal in the upper part of this figure.}
    \label{fig_nikhef_knocking}
  \end{center}
\end{figure}

The result of the first measurement with the acoustic sensor is shown in Fig.~\ref{fig_nikhef_spectrum}.
Five PMT bases (without PMT) were connected and powered for this measurement.
It shows the noise spectrum with an essentially flat response up to about $125\,\textrm{kHz}$.
Afterwards the bandpass filter suppresses the noise as expected.
The response of the acoustic sensor to an acoustic signal is shown in Fig.~\ref{fig_nikhef_knocking}.
The acoustic signal was generated by tapping the glass sphere with a fingertip in the same readout configuration as before.
This test was done to verify the correct operation of the sensor and provides no physical content.
The prototype module will be tested in situ in the context of the ANTARES \cite{ANTARES} ``Instrumentation Line''.
Further tests are planned to optimise the design.

\section{Conclusions and outlook}

The experimental results for the different prototype OAMs show mostly the expected behaviour.
The experience made with the Acoustic Modules (AMs) in AMADEUS \cite{amadeus}, OAMs without PMT(s), and their satisfying accuracy for their positioning inside ANTARES make the OAMs a promising option for the acoustic positioning system in KM3NeT.
The elevated noise level measured in air mainly result from insufficient acoustic shielding against the noisy laboratory environment as well as from electromagnetic interference.
The electromagnetic interferences will require small modifications to the acoustic sensor, e.g.~better shielding or modified signal transfer, and a better integration/adaptation to the optical part of the module.
The OAMs in the multi-PMT design allow for using equal-design modules and thus a simplified construction of KM3NeT.
The piezo sensors used for the OAMs are not solely designed for the acoustic positioning system as their sensitivity and the whole data readout system allows for a wider range of applications like monitoring of the deep sea acoustic environment and studies towards acoustic particle detection based on the thermo-acoustic model \cite{thermo_acoustic_model}.
The ability to extend the scope of a conventional water Cherenkov telescope towards the highest energies and to a variety of multidisciplinary subjects demonstrates the potential of the described sensor type.

\section*{Acknowledgments}
The research leading to these results has received funding from the European Community Seventh Framework Programme under grant agreement n.~212525 and from the Sixth Framework Programme under contract n.~011937, as well as by the German Federal Ministry for Education and Research (BMBF) under grants 05A08WE1 and 05A11WE1.

\end{document}